\documentclass[a4paper,twocolumn,11pt,accepted=2021-03-08]{quantumarticle}
\pdfoutput=1
\usepackage[utf8]{inputenc}
\usepackage[english]{babel}
\usepackage[T1]{fontenc}
\usepackage{amsmath}
\usepackage{hyperref}
\usepackage[numbers,sort&compress]{natbib}
\usepackage{url}

\usepackage{tikz}
\usepackage{lipsum}

\usepackage{amssymb,amsmath,amsfonts,amsbsy,amsthm,bm}
\usepackage{dsfont}
\usepackage{bbold}

\usepackage{graphicx}

\def\<{\langle}
\def\>{\rangle}
\def\dag{\dagger}

\newcommand{\tr}{\text{Tr}}
\newcommand{\ket}[1]{| #1 \rangle}
\newcommand{\bra}[1]{\langle #1 |}

\newcommand{\var}{\text{Var}}
\newcommand{\cov}{\text{Cov}}

\begin{document}

\title{Estimating expectation values using approximate quantum states}

\author{Marco Paini}
\email{mpaini@rigetti.com}
\affiliation{Rigetti Computing, 138 Holborn, London, EC1N 2SW, UK.}

\author{Amir Kalev}
\email{amirk@isi.edu}
\affiliation{Information Sciences Institute, University of Southern California, Arlington, VA 22203, USA.}

\author{Dan Padilha}
\affiliation{Rigetti Computing, 138 Holborn, London, EC1N 2SW, UK.}

\author{Brendan Ruck}
\affiliation{Rigetti Computing, 2919 Seventh St, Berkeley, CA 94710, USA.}
\maketitle

\begin{abstract}
We introduce an approximate description of an $N$--qubit state, which contains sufficient information to estimate the expectation value of any observable to a precision that is upper bounded by the ratio of a suitably--defined  seminorm of the observable to the square root of the number of the system's identical preparations $M$, with no explicit dependence on $N$. We describe an operational procedure for constructing  the approximate description of the state that requires, besides the quantum state preparation, only single--qubit rotations followed by single--qubit measurements. We show that following this  procedure, the cardinality of the resulting description of the state grows as $3MN$.  We test the proposed method on Rigetti's quantum processor unit with 12, 16 and 25 qubits for random states and random observables, and find an excellent agreement with the theory, despite experimental errors. 
\end{abstract}

\section{Introduction}
Quantum density operators represent our knowledge of the state of quantum systems and give us a way to calculate expectation values and predict experimental results via the Born rule. Our knowledge of a system's density operator is equivalent to our ability to calculate the expectation value of any observable of the system. However, for an $N$--body system, reconstructing the system's density operator requires the knowledge of exponentially many numbers in $N$. This in turn translates into practical difficulties in estimating and storing density operators, in the form of density matrices, even for systems composed of just a few tens of  qubits.

Over the last few years, there has been a surge of interest in devising protocols to reduce the resources required to estimate and classically store quantum states~\cite{Vidal2003Efficient,Aaronson2007learnability,Cramer2010Efficient,Gross2010Quantum,Kalev2015Quantum,Gao2017Efficient,Torlai2018Neural,Aaronson2018Shadow,Gosset2019compressed}. 
There have been numerous approaches to reduce the cost of performing quantum state tomography by assuming that the state of the system has a specific  structure. These approaches include, for example:  matrix product states tomography~\cite{Cramer2010Efficient}, where the state is assumed to have an efficient matrix product state representation as a matrix; neural network tomography~\cite{Gao2017Efficient,Torlai2018Neural}, where the state is assumed to have an efficient deep neural network representation; and compressed sensing tomography~\cite{Gross2010Quantum,Kalev2015Quantum}, where the state is assumed to have a low-rank density matrix representation. The theory of Probably Approximately Correct (PAC) learning was utilised in~\cite{Aaronson2007learnability} to show that if, instead of full tomography, we are interested in approximating with high probability the expectation values of observables drawn from a distribution, then only polynomially--many samples in $N$ of the observables' expectation values are required. It was later also shown~\cite{Aaronson2018Shadow} that estimating with high probability the expectation values of a set of observables can be achieved  using only polynomially--many copies in $N$ of the system. As for storage, it is well established~\cite{Vidal2003Efficient} that quantum states that can be represented as low bond--dimension matrix product states can be efficiently stored on classical computers. Recently it was shown~\cite{Gosset2019compressed} that a general quantum state of $N$ qubits can be approximately described using its inner product with an order of $\sqrt{2^N}$ stabiliser states. This result thus provides a polynomial improvement for classically storing quantum states over a brute--force approach of storing density matrices. In Sec.~\ref{Related work} we discuss  these and other works in connection to the results presented in this paper.

The growth of the number of copies of the system in $N$ for estimation and the considerable requirements in $N$ for classically storing the information associated with a generic quantum state lead to the question of whether a protocol that significantly reduces or eliminates the dependence on $N$ is admissible. In this work, we give an affirmative answer to the question. Instead of focusing on the estimation and storage of the density matrix, we start by considering the equivalent problem of calculating the expectation value of any observable of the system up to a certain precision. In this article we show that, for a system of $N$ qubits, the expectation value of any observable can be calculated with arbitrary precision with a number of copies of the system independent of $N$, but dependent on a properly--defined seminorm of the observable. The information obtained through the measurements on the copies of the system will be used to form an {\it approximate description of the quantum state} -- a description with a cardinality that grows linearly in $N$ and thus can be constructed and stored for large systems. The method for estimating expectation values,  and the exact definition of an observable seminorm and of the storage approximate quantum state will be given in the next section.

\section{Expectation value estimation and the approximate quantum state} \label{approximate description}
We will consider specifically a system of $N$ qubits. However, the considerations that will be made are applicable in general to other systems, with finite and infinite--dimensional Hilbert spaces, since they are based on the general theory of irreducible representations of groups which provide explicit forms for a basis in the space of linear bounded operators defined on a Hilbert space. We refer the reader to Refs. ~\cite{Paini2000Quantum,DAriano2003Spin} where these considerations have been developed.

We start by presenting the method to estimate the expectation value of any observable and notion of the approximate state for a single--qubit system. We  then generalise these ideas  for the more general $N$--qubit case. 

\subsection{One qubit} \label{tomo 1 qubit}
A single--qubit density operator can be written as~\cite{Paini2000Quantum,DAriano2003Spin}
\begin{align}
\rho=\int_{\Sigma}\frac{d\vec n}{4\pi}\sum_{m=\pm1}p(m,\vec n)K_1(m,\vec n),				\label{tomo 1-qubit}
\end{align}
with the unitary spherical surface ${\Sigma}$ as integration domain, $\vec n=(\cos \varphi\sin\vartheta,\sin\varphi\sin\vartheta,\cos\vartheta)$, with $\vartheta\in[0,\pi],\varphi\in[0,2\pi)$, is a unit vector in $\mathds{R}^3$, $p(m,\vec n)$ is the probability of obtaining the result $m\in\{-1,+1\}$ when measuring the qubit in the eigenbasis of $\vec \sigma\cdot\vec n$, where $\vec \sigma=(\sigma_x,\sigma_y,\sigma_z)$ is a vector of the Pauli operators, and  $K_1(m,\vec n)$ is the 1--qubit kernel operator given by
\begin{align}
K_1(m,\vec n)=\frac{1}{2}(\mathbb{1}+ 3\,m\,\vec{\sigma}\cdot\vec{n}), 	\label{kernel}
\end{align}
see App.~\ref{app 1 qubit} for more details. 

Let us define a single--qubit (unbiased) {\it estimator function}
\begin{align}\label{1-qubit estimator func}
{R}_1[\;\cdot\;](m,\vec n)=\tr\Bigl[\;\cdot\; K_1(m,\vec n)\Bigr],
\end{align}
where the $\cdot$ represents the argument of the function ${R}_1$, i.e. a single--qubit  observable (for a lighter notation we sometime omit the explicit dependence of the estimator on $(m,\vec n)$). In particular, the expectation value of the Pauli operator $\sigma_i$, with $i=0,1,2,3$ and $\sigma_0:=\mathbb{1},$  $\sigma_1:=\sigma_x,\,\sigma_2:=\sigma_y,$ and $\sigma_3:=\sigma_z$, can be written in terms of ${R}_1[\sigma_i]$ as 
\begin{align}
\langle\sigma_i\rangle=\int_{\Sigma}\frac{d\vec n}{4\pi}\sum_{m=\pm1}p(m,\vec n) {R}_1[\sigma_i](m,\vec n).											\label{<sigma>}
\end{align}
Calculating the estimator of the Pauli operators  we arrive at
\begin{align}
{R}_1[\mathbb{1}](m,\vec n)&=1, 						\label{identity estimator}\\ 
{R}_1[\sigma_\alpha](m,\vec n)&=3\,m\,n_\alpha,\,\,\text{for}\,\, \alpha=x,y,z.		\label{pauli estimator}
\end{align}
The estimator ${R}_1[\sigma_\alpha](m,\vec n)$  is a random variable due to its  dependence on the stochastic measurement outcome $m$ and the random unit vector $\vec{n}$.  Hereafter, unless noted otherwise, we use the subscript $\alpha$ to index $x,y,z$ and  subscript $i$ to index $0,1,2,3$. In App.~\ref{app 1 qubit} we show that the variance  of the estimator ${R}_1[\sigma_\alpha]$ is upper bounded by $3$,
\begin{align}
\var({R}_1[\sigma_\alpha](m,\vec n))&\leqslant3.			\label{pauli variance}
\end{align}
Since ${R}_1[\mathbb{1}](m,\vec n)$ is independent of $m$ and $\vec n$, it is deterministic and thus has a null variance.

Equation~\eqref{<sigma>} implies that we can construct an {\it  estimate} of $\langle\sigma_i\rangle$ using the following Monte Carlo sampling procedure:
\begin{enumerate}
\item Sample a point  $\vec n$ uniformly from the surface of a unit sphere, 
\item prepare the qubit in the state $\rho$ and measure it in the eigenbasis of $\vec \sigma \cdot\vec n$, 
\item record the outcome $m\in\{-1,1\}$,
\item repeat steps 1--3 $M$ times. 
\end{enumerate}
The estimate of  $\langle\sigma_i\rangle$, which we denote by $\langle\sigma_i\rangle_M$, is simply given by the average of corresponding estimator ${R}_1[\sigma_i](m,\vec n)$ over the data set:
\begin{align}
\langle\sigma_i\rangle_M=\frac1{M}\sum_{j=1}^M {R}_1[\sigma_i](m_j,\vec n_j) ,															\label{<R1sigma>}
\end{align}
where $\vec n_j=(\cos \varphi_j\sin\vartheta_j,\sin\varphi_j\sin\vartheta_j,\cos\vartheta_j)$ is the unit vector sampled at the $j$th iteration and $m_j$ is the corresponding measurement outcome. Note that the estimate $\langle{\mathbb{1}}\rangle_M=1$ for any value of $M$ and it has  zero variance. Moreover, by construction, in the limit $M{\to}\infty$, $\langle\sigma_\alpha\rangle_M{\to}\langle\sigma_\alpha\rangle$, for $\alpha=x,y,z$. The value of $M$ determines the precision of the estimation. For $M$ sufficiently large,  $\langle\sigma_\alpha\rangle_M$ has a normal statistical error, which due to Eq.~\eqref{pauli variance} is smaller than or equal to $\sqrt{3/M}$.  

The measurement in the eigenbasis of $\vec \sigma \cdot\vec n$, at step~2 above, can be implemented by first applying a unitary operator $U$, that rotates this basis to the eigenbasis of $\sigma_z$, and then measuring in the eigenbasis $\sigma_z$ (the computational basis). This unitary operator is given by $U=e^{i\frac{\vartheta}{2}\vec\sigma\cdot\vec n_\perp}$, where $\vec n_\perp=(-\sin\varphi,\cos\varphi,0)$.  We call the vector $s_j=(m_j,\vartheta_j,\varphi_j)$ the $j$-th  {\it snapshot}  of the qubit system. We refer to the collection of the $M$ snapshots, ${\cal S}=\{s_j\}_{j=1}^M$, that we obtain following the sampling procedure above,  as the {\it approximate quantum state}, as it enables us to estimate the expectation value of any single--qubit observable to a certain precision as we now describe. 

Let $O$ be a single--qubit observable, expanded  in the Pauli basis $\{\sigma_i\}_{i=0}^3$,
\begin{align}
O=\sum_{i=0}^{3}a_i\,\sigma_i. 															\label{O}
\end{align}
Given the estimated expectation values of $\{\sigma_i\}_{i=0}^3$, by linearity of the estimator function~\eqref{1-qubit estimator func}, we can construct an estimate of $\< O\>$ as
\begin{align}
\langle{O}\rangle_M=\sum_{i=0}^{3}a_i\,\langle\sigma_i\rangle_M.											\label{R linear}
\end{align}
We prove in App. \ref{app 1 qubit} that for $M$ sufficiently large, $\langle{O}\rangle_M$ has a normal statistical error smaller than or equal to $\Vert O\Vert/{\sqrt{M}}$, that is
\begin{align}
\text{Std}(\langle{O}\rangle_M)\leqslant \frac{\Vert O\Vert}{\sqrt{M}}\,,									\label{O variance}
\end{align}
where  Std stands for `standard deviation' and
\begin{align}
\Vert O\Vert=\sqrt{3\sum_{i=1}^{3}a_i^2}							\label{norm 1 qubit}
\end{align}
is a seminorm of $O$ that depends on its expansion in the Pauli operator basis.

\subsection{\boldmath{$N$} qubits}
The method developed for one qubit in the previous subsection is easily generalised to $N$ qubits, in the assumptions that the qubits are distinguishable and that a qubit--specific measurement capability is available. The space of the $N$--qubit system is the tensor product of the single--qubit spaces and thus we can write a general $N$--qubit density operator as~\cite{Paini2000Quantum,DAriano2003Spin}
\begin{align}
\rho=&\bigg({\frac{1}{(4\pi)^N}\prod_{k=1}^N \int_{{\Sigma}_k}{d{\vec n}_k}\sum_{m_{k}}}\bigg)p(\{m,\vec n\})K(\{m,\vec n\}),				\label{tomo N-qubit}
\end{align}
where $\{m,\vec n\}$ is a shorthand for $(m_{1},{\vec n}_1,\ldots,m_{N},{\vec n}_N)$, the $N$--qubit kernel operator $K(\{m,\vec n\})$ equals to the tensor product of single--qubit kernels
\begin{align}
K(\{m,\vec n\})=\bigotimes_{k=1}^N K_1(m_{k},{\vec n}_k),	\label{kernel N-qubit}
\end{align}
and $p(\{m,\vec n\})$ represents the joint probability of finding the outcome $(m_{1},\ldots,m_{N})\in\{-1,1\}^{N}$ when measuring the qubits in the product eigenbasis of ${\vec \sigma}_1\cdot{\vec n}_1\otimes\cdots\otimes{\vec \sigma}_N\cdot{\vec n}_N$. 
Similarly to the single qubit case, this measurement can be implemented with single--qubit rotations followed by a measurement in the computational basis on each qubit.

We define the $N$--qubit  (unbiased) estimator function
\begin{align}\label{N-qubit estimator func}
{R}[\;\cdot\;](\{m,\vec n\})&= \tr\Bigl[\;\cdot\; K(\{m,\vec n\})\Bigr].
\end{align}
As a result of the tensor product structure of the kernel operator,  ${R}[P_i]$, where $P_i=\sigma_{i_1}\otimes\cdots\otimes\sigma_{i_N}$,  $i=(i_1,\ldots\,i_N)$, is a monomial of Pauli operators (a `Pauli monomial' or a `Pauli string' for short), and hereafter $i_k=0,\ldots,3$, is simply the product of the corresponding single--qubit random variables,
\begin{align}
{R}[P_i](\{m,\vec n\})&=\prod_{k=1}^N {R}_1[\sigma_{i_k}](m_k,\vec n_k),
\label{R_N}
\end{align}
where $R_1[\sigma_{i_k}](m_{k},\vec{n}_{k})$ is given in Eqs.~\eqref{identity estimator} and \eqref{pauli estimator}.

\begin{figure}[t!]
\centering
\includegraphics[width=1\columnwidth]{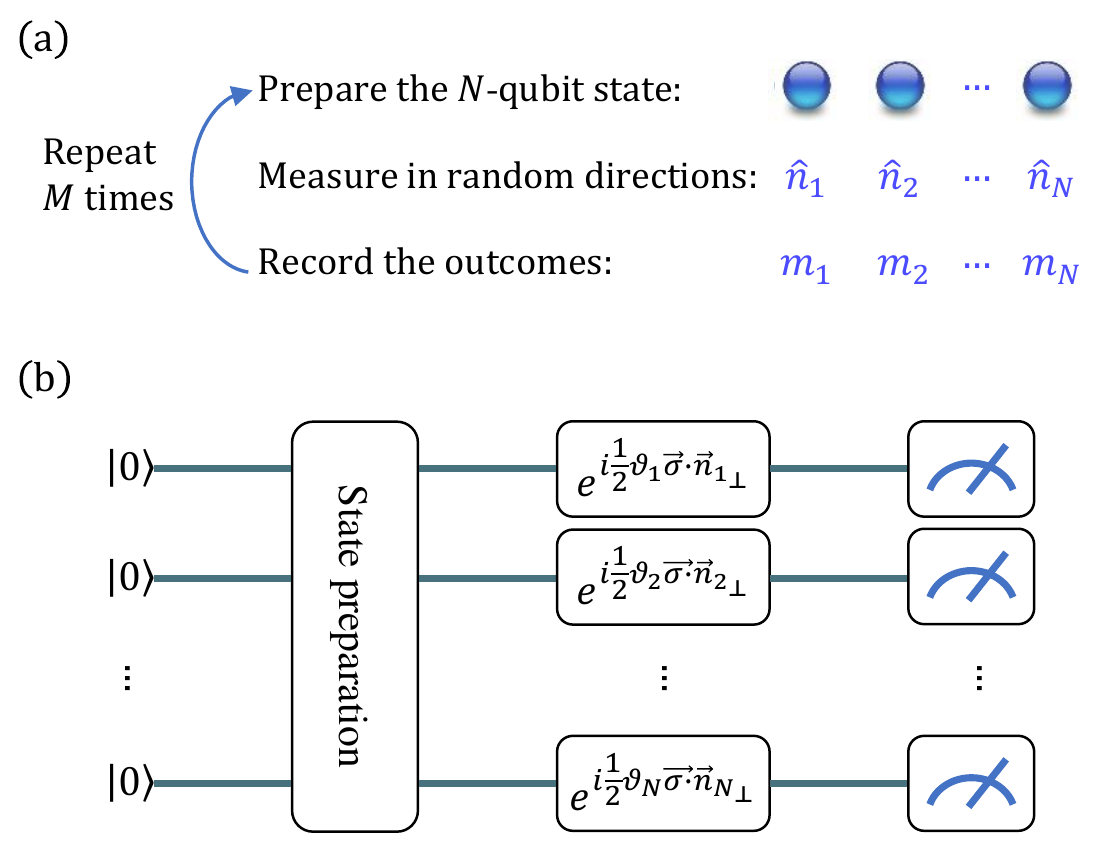}
\caption{{\bf Obtaining the approximate quantum state.} (a) To construct the approximate description of an $N$--qubit state we repeat the following steps $M$ times: (1) prepare a copy of the $N$--qubit state, (2) measure each qubit in a random direction ${\vec n}_j=(\cos \varphi_j\sin\vartheta_j,\sin\varphi_j\sin\vartheta_j,\cos\vartheta_j)$, (3) record the measurement outcome  of each qubit.  (b) A circuit realization of the protocol. The single--qubit rotations are defined by $\vartheta_j$ and $\vec n_{j_{\perp}}=(-\sin\varphi_j,\cos\varphi_j,0)$, $j=1,\ldots,N$.
}\label{fig:circuit}
\end{figure}

Similarly to the single qubit case of the previous section, we can write the expectation value of $P_i$ with respect to the state of the system in terms of ${R}[P_i]$ as
\begin{align}\label{<sigmaNqubit>}
\langle P_i\rangle=&\bigg({\frac{1}{(4\pi)^N}\prod_{k=1}^N \int_{{\Sigma}_k}{d{\vec n}_k}\sum_{m_{k}}}\bigg)\\\nonumber&\times p(\{m,\vec n\}) {R}[P_i](\{m,\vec n\}).
\end{align}
Eq.~\eqref{<sigmaNqubit>} implies that an estimate $\langle P_i\rangle_M$ of $\langle P_i\rangle$ can be calculated by performing the single--qubit tomographic procedure introduced in the above section, in parallel on all of the $N$ qubits, as in Fig.~\ref{fig:circuit}.  We call the vector $s_j=({m_1}_j,{{}{\vec n}_1}_j,\ldots,{m_N}_j,{{}{\vec n}_N}_j)$ the $j$th {\it snapshot} of a system of $N$ qubits, and we refer to a set $\cal S$ of  such $M$ independent snapshots as the {\it approximate state} of the $N$--qubit system.

The use of single--qubit Haar--random basis measurements was recently proposed, and experimentally implemented, to probe certain properties of an $N$--qubit system, such as purity~\cite{Brydges2019Probing,Elben2020Many} and the Frobenius inner product of two states~\cite{Elben2020Cross}. In light of these results, as we now show, the measurement protocol we introduced above is a universal protocol for estimating the expectation value of any $N$--qubit observable. The protocol is efficient in terms of time and space resources, as long as the observable admits an efficient representation in the Pauli basis and has a small seminorm (e.g. not growing exponentially with $N$).  

To start with, we can use the approximate state  $\cal S$ to calculate an  estimate $\langle P_i\rangle_M$ of  $\langle{P_i}\rangle$ as
\begin{align}
\langle P_i\rangle_M&=\frac1{M}\sum_{j=1}^M {R}[P_i](\{m,\vec n\})\nonumber\\&= \frac1{M}\sum_{j=1}^M\prod_{k=1}^N R_1[\sigma_{i_k}]({m_k}_j,{{}{\vec n}_k}_j),					\label{general string estimator}
\end{align}
where the first equality follows from Eq.~\eqref{R_N}.  We show in App.~\ref{app N qubits} that for $M$ sufficiently large,  $\langle P_i\rangle_M$ ($i\neq 0_v\doteq (0,\ldots,0)$)  has a normal statistical error that is smaller than or equal to $\sqrt{3^{r_i}/M}$, where $r_{i}$ represents the weight of $P_i$, i.e., the number of qubits on which it acts non--trivially. Note that the estimated expectation value for the $N$--qubit identity operator $\langle \mathbb{1}^{\otimes N}\rangle_M=1$ and it has zero variance.

With the ability to efficiently calculate the estimated expectation value of any given Pauli monomial operator, we can estimate the expectation value of $N$--qubit observables.  Let $O$ be an $N$--qubit observable such that 
\begin{align}
O=\sum_{i} \,a_i\,P_i\,,								\label{general O}
\end{align}
From linearity of the estimator function~\eqref{N-qubit estimator func} and from Eq.~\eqref{general string estimator}, the estimated expectation value of $O$ is given by
\begin{align}
\langle{O}\rangle_M&=  \sum_ia_i \langle{P_i}\rangle_M\nonumber\\&=\frac1{M}\sum_{j=1}^M \sum_ia_i\prod_{k=1}^NR_1[\sigma_{i_k}]({m_k}_j,{{}{\vec n}_k}_j).						\label{general estimator}
\end{align}
Similar to $\langle{P_i}\rangle_M$, $\langle{O}\rangle_M$ is an unbiased estimate, i.e., $\langle{O}\rangle_M{\to}\langle{O}\rangle$ in the limit  $M{\to}\infty$.  Practically, for the efficient calculation of \eqref{general estimator}, the sum on $i$ should have at most a polynomial number of terms in $N$. 

While we  assumed in Eq.~\eqref{general O} that the observable admits an efficient representation in terms of Pauli monomials,  as a result of the tensor product structure of the kernel operator, Eq.~\eqref{kernel N-qubit}, our procedure is efficient to calculate the expectation value of any operator of the form 
\begin{align}
O=\sum_{i} \,a_i\,O_{i_1}\otimes O_{i_2}\cdots\otimes O_{i_N}\,,									\label{more general O}
\end{align}
where $O_{i_k}$ is a single-qubit operator (on the $k$th qubit space). If this is the case then, 
\begin{align}
\langle O\rangle_M=\frac1{M}\sum_i a_i \sum_{j=1}^M\prod_{k=1}^N R_1[O_{i_k}].						\label{more general estimator}
\end{align}
Note that any Pauli monomial $P_i$ is a special case of $O_{i_1}\otimes O_{i_2}\cdots \otimes O_{i_k}$. 

We show in App.~\ref{app N qubits} that the standard deviation of $\langle{O}\rangle_M$ has the upper bound
\begin{align}
{\rm Std}(\langle{O}\rangle_M)\leqslant\frac{{\Vert O\Vert}}{\sqrt M},						\label{general variance}
\end{align}
with the seminorm of $O$ defined as
\begin{align}
\Vert O\Vert=\sqrt{\sum_{i,j (\neq 0_v)}3^{r_{ij}}\Delta_{ij}|a_i||a_j|},
 										\label{norm N qubits}
\end{align}
where the sum is extended to all values of $i$ and $j$ except $0_v\doteq(0,0,\ldots,0)$, $r_{ij}$ is the number of indices $k$ for which $i_k\neq 0 \wedge  j_k\neq 0$, $\Delta_{ij}=0$ if there exists a $k$ such that $i_k\neq 0 \wedge  j_k\neq 0 \wedge i_k\neq j_k$ and $\Delta_{ij}=1$ otherwise. Eq.~\eqref{norm N qubits} reduces to Eq.~\eqref{norm 1 qubit} when $O$ is an observable on a single--qubit space.

Importantly, in this procedure, the expectation values estimation error has no global dependence on the system size $N$. The error only depends on the observable seminorm and the number of snapshots. One could argue that the presence in the seminorm of $3^{r_{ij}}$ reflects an exponential growth of the estimates' variance in $N$, as, for example, for an observable given by the tensor product of $N$ Pauli operators, which has seminorm ${3}^{N/2}$. However, the seminorm is a property of the observable and not of the dimension of the space: an $N$--qubit observable $O$ and its extension to a larger $(N{+}L)$--qubit space, $O\otimes\mathbb{1}^{\otimes L}$, have the same seminorm~\eqref{norm N qubits}, independent of $L$. Irrespective of $N$, because of Eq.~(\ref{general variance}), the expectation values of all observables with unit seminorm can be obtained with statistical error bounded by $1/\sqrt{M}$. Any observable  with seminorm different from zero can be obtained from an observable with unit seminorm through multiplication by a real number, representing in fact the observable's seminorm. For a given number of identical preparations and measurements, the statistical error in the expectation value of the generic observable has a bound only dependent on how ``large'' the observable is. In App.~\ref{app N qubits}, we also explain how the  seminorm ${\Vert O\Vert}_2$, defined by
\begin{align}
{\Vert O\Vert}_2=\sqrt{\sum_{i\neq 0_v}3^{r_{i}}\,a_i^2}, 						\label{original norm}
\end{align}
divided by $\sqrt{M}$, with $r_i$ representing the weight of $P_i$,  is generally a good approximation for the standard deviation of $\langle{O}\rangle_M$. This result may be important for applications of the method since ${\Vert O\Vert}_2$ can be significantly smaller than $\Vert O\Vert$ and since, for a chosen statistical error, the number of identical preparations required is inversely proportional to the variance of $\langle{O}\rangle_M$.  See App.~\ref{app:pps} for further details.

We close this part with a remark on the notion of the approximate  quantum state, $\cal S$.  We have shown that the approximate description of the state,  as we have defined it, comes  with an explicit operational procedure to estimate the expectation value of any generic observable with a rigorous bound on the estimation statistical error. Therefore, the approximate state together with the rule~\eqref{general estimator} provides a description of a physical system in terms of statistical inferences based on observed experimental data and, in principle, could be assumed as primitive concepts instead, or, at least, as not requiring the existence of density operators and the Born rule.

\section{Experimental results} \label{experimental results}
As a proof--of--concept, we have tested the theory on Rigetti Computing quantum processor unit (QPU), Aspen7-28Q-A,  with three systems, involving $N=12, 16$ and $25$ qubits, respectively. In each experiment the system is prepared in a random state. The state preparation circuit is composed of two layers of gates applied to the  computational state $|0\rangle^{\otimes N}$. In the first layer, the gates are chosen at random from a flat distribution over the single--qubit gate set $\{X,Y, Z, H, S, T\}$ and applied to $N/2$ qubits, chosen at random. While in the second layer the two--qubit gate $XY(\alpha)=e^{-i\alpha(X\otimes X+Y\otimes Y)}$ is applied to $50\%$ of pairs of qubits, chosen at random, with different values of $\alpha$ sampled uniformly from $[0,2\pi)$. This gate composition results in shallow quantum circuits with $N/2$ random single--qubit gates and $N/4$  random two--qubit gates, to ensure that the experimentally prepared state has a high fidelity with the target state. 

\begin{figure*}[t!]
\begin{minipage}[b]{.5\linewidth}
\centering
\includegraphics[width=0.95\linewidth]{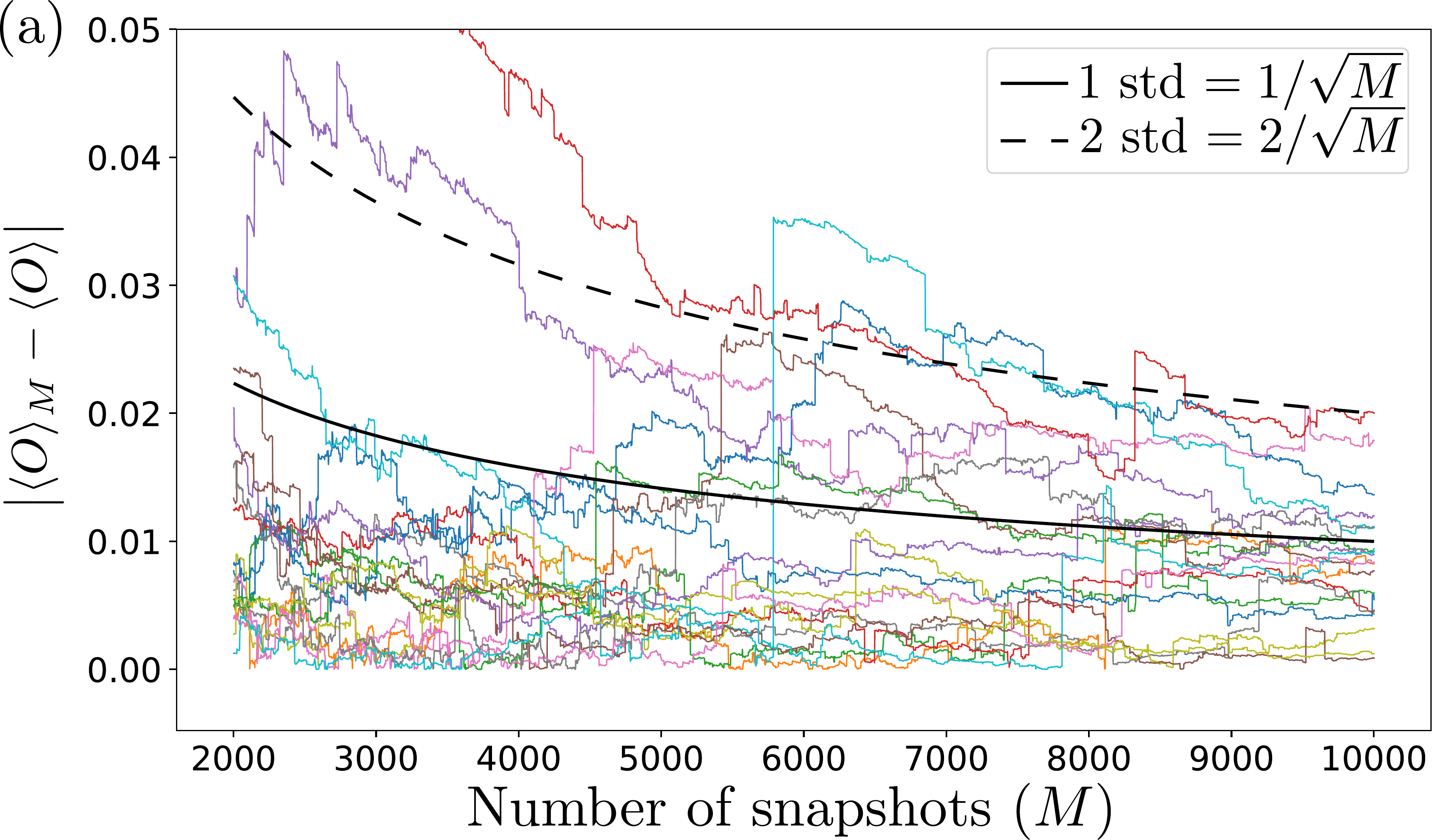}
\end{minipage}%
\begin{minipage}[b]{.5\linewidth}
\centering\includegraphics[width=0.95\linewidth]{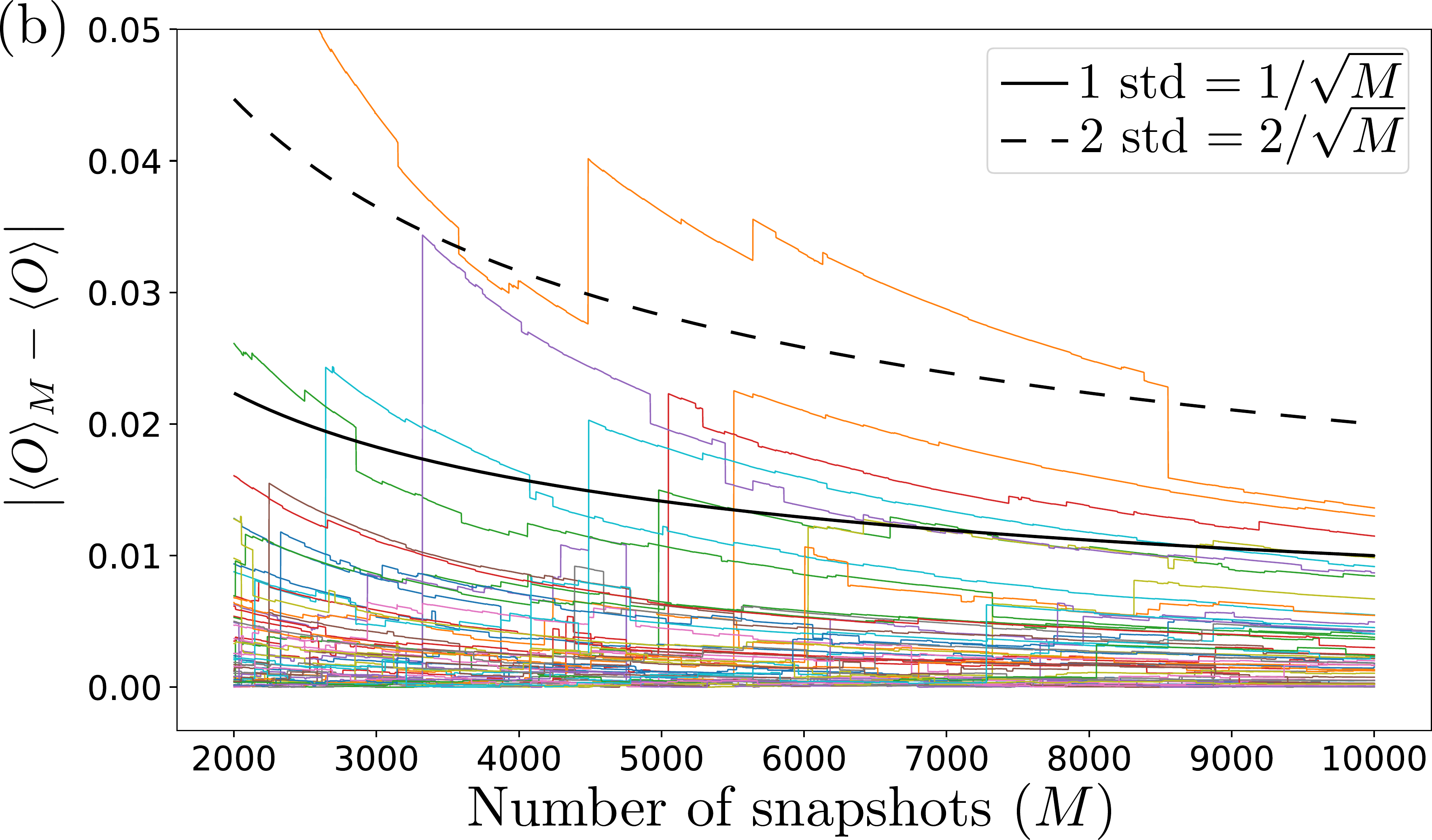}
\end{minipage}
\caption{The error in the estimated expectation value of (a) 20 random observable and (b) 20 random projectors onto computational basis states in a 25--qubit system as a function of the number of snapshots.}\label{fig:Q25}
\end{figure*}

For each system, an approximate description of the prepared state was obtained using $M=10^4$ snapshots (to check the results to $0.01$ accuracy).  The approximate state was then used to estimate the expectation value of two sets of observables: (1) 20 random observables, (2) 20 random projectors onto computational basis states (we have selected 20 for the number of states  so that it would be visually `easy' to  check the results in Fig.~\ref{fig:Q25} against the second standard deviation given it is at $\sim$95\%). Each random observable in the first set is a linear combination of 20 Pauli monomials. Each Pauli monomial  is constructed by randomly sampling from the set $\{\sigma_i\}_{i=0}^{3}$ for each qubit, whereas the linear combination coefficients (the $a_i$'s) are sampled uniformly from the $(0,1]$. The projectors onto computational basis states were sampled uniformly from the set $\{0,1\}^N$. For simplicity in the analysis, the 20 random observables are normalised to have a unit seminorm (the projectors already have unit seminorm with probability exponentially close to 1 as a function of $N$, see App.~\ref{app:pps}).

Fig.~\ref{fig:Q25} shows the results for the 25--qubit experiment. In this figure, we compare the estimated expectation values of random observables and of random computational basis projectors  to their expectation values with respect to the 25--qubit target state. The results are plotted as a function of the number of snapshots.  Table~\ref{tbl:std} summarises the experimental results of the $12$--, $16$-- and  $25$--qubit systems for $10^4$ snapshots. The figure and the table show that the experimental results generally mirror the theory, with the fraction of estimates within one and two standard deviations from their true values not far from $68\%$ and $95\%$ respectively. Other than statistical considerations, the cases with a fraction of the estimates greater than the expected percentages from a normal distribution can be explained in terms of how tightly the observables' seminorms bound the standard deviation of the estimators, in particular for the case of the projectors. 
\begin{table}[h!]
\begin{center}
\flushleft{(a) Random observables}
\begin{tabular}{ |c||c|c|c|}
  \hline
  Number of qubits&~12&~16&~25\\
  \hline
  Estimation within 1 Std  &~$60\%$ &~$65\%$&~$70\%$ \\
Estimation within 2 Std's  &~$93\%$&~$100\%$&~$95\%$\\
    \hline
 \end{tabular}\\
\flushleft{(b) Random projectors}
\begin{tabular}{ |c||c|c|c|}
  \hline
  Number of qubits&~12&~16&~25\\
  \hline
  Estimation within 1 Std  &~$70\%$ &~$90\%$&~$85\%$ \\
Estimation within 2 Std's  &~$95\%$&~$95\%$&~$100\%$\\
    \hline
 \end{tabular}
\end{center}\vspace{-0.3cm}
\caption{\label{tbl:std} The fraction of the estimated expectation values, of (a) random observables and (b) random projectors onto  computational basis states, that are within one and two standard deviations from their true expectation values for $10^4$ snapshots.}
\end{table} 

To complete the picture, including an explanation for the cases of a lower fraction of the estimates than predicted, we need to start examining the effects of hardware errors. It is worth noting that  we obtain  agreement with the theory despite an order of $p_{\rm err}=5\%$ (averaged) readout error per qubit, which amounts to flipping the measurement outcome from $+1$ to $-1$ and vice versa.  The resilience  of our protocol to single--qubit readout errors can be attributed to the randomness of the measurement bases, as was discussed in~\cite{Vermersch2019Probing}. Moreover, the following considerations provide additional insights in the robustness of the estimation protocol to readout errors.

Firstly, let us consider the estimation of the expectation value of single--qubit observable. Let $\cal S$ be the approximate state description composed of $M$ snapshots. Suppose that in $M_{\rm err}$ of the snapshots we obtained the erroneous measurement outcome, $M_{\rm err}/M\approx p_{\rm err}$. The estimation of the expectation value of any single--qubit Pauli operator  $\sigma_\alpha$ ($\alpha=x,y,z$) is
 \begin{align}
\langle{\sigma_\alpha}\rangle_M&=\frac1{M}\sum_{j=1}^{M} 3m_j{n_\alpha}_j\nonumber\\
&=\frac{M-M_{\rm err}}{M}\Big(\frac1{M-M_{\rm err}}\sum_{j=1}^{M-M_{\rm err}} 3m_j{n_\alpha}_j\Big)\nonumber\\&+\frac{M_{\rm err}}{M}\Big(\frac1{M_{\rm err}}\sum_{j=1}^{M_{\rm err}} 3m_j^{\rm err}{n_\alpha}_j\Big),
\end{align}
where $m_j^{\rm err}$ is the erroneous outcome. Since $m_j^{\rm err}=-m_j$ we arrive at
 \begin{align}
\langle{\sigma_\alpha}\rangle_M&=\frac{M-M_{\rm err}}{M}\Big(\frac1{M-M_{\rm err}}\sum_{j=1}^{M-M_{\rm err}} 3m_j{n_\alpha}_j\Big)\nonumber\\&-\frac{M_{\rm err}}{M}\Big(\frac1{M_{\rm err}}\sum_{j=1}^{M_{\rm err}} 3m_j{n_\alpha}_j\Big),
\end{align}
which, for sufficiently large $M$ and $M_{\rm err}$, can be approximated by
 \begin{align}
\langle{\sigma_\alpha}\rangle_M&\approx(1-p_{\rm err})\<{\sigma_\alpha}\>-p_{\rm err}\<{\sigma_\alpha}\>\nonumber\\&=(1-2p_{\rm err})\<{\sigma_\alpha}\>.
\end{align}
Therefore, for a single--qubit observable $O=\sum_i a_i\sigma_i$ we arrive at
 \begin{align}
\langle{O}\rangle_M&\approx a_0\mathbb{1}+\sum_{i=1}^3a_i(1-2p_{\rm err})\<{\sigma_i}\>.
\end{align} 
Thus, the readout error introduces an attenuator $(1-2p_{\rm err})$ to each coefficient different than $a_0$ in the expectation value.
 
Moving to the $N$--qubit case, we first note that only odd numbers of readout errors per snapshot contribute to the estimation error of an $N$--qubit expectation value. This is because  the estimated expectation value is calculated via  a product of the measurement outcomes which take values $\pm1$, c.f. Eq.~\eqref{general string estimator}. Therefore, if a given snapshot includes readout errors on an even number of qubits, there are even number of erroneous outcomes $m_{j}^{\rm err}$ such that $m_{j}^{\rm err}= -m_j$, and the estimated expectation value remains unaffected by these errors.  Let $p_{\rm odd}(r)$ be the probability that in a given snapshot there is an odd number of readout errors on $r$ qubits. Given a readout error $p_{\rm err}$ per qubit,  
\begin{align}
p_{\rm odd}(r)&=\sum_{k~\rm {odd}}{r\choose{k}}p_{\rm err}^k(1-p_{\rm err})^{r-k}\nonumber\\&= \frac{1}{2}\Big(1-(1-2 p_{\rm err})^r\Big).
\end{align}
Similarly to the single qubit analysis, one can then show that the expectation value of a Pauli monomial $P_i$ with weight $r_i$ is 
\begin{align}
\langle{P_i}\rangle_M&\approx(1- 2p_{\rm odd}(r_i))\<{P_i}\>=(1- 2p_{\rm err})^{r_i}\<{P_i}\>.
\end{align}
We therefore obtain that the estimated expectation value of an $N$--qubit observable $O=\sum_i \,a_i\,P_i$ is given by
\begin{align}\label{eq:readout_err}
\langle{O}\rangle_M&\approx\sum_{i}a_i(1-2 p_{\rm err})^{r_i}\<{P_i}\>.
\end{align}
This equation indicates that the readout error does not depend on $N$. It introduces an attenuator $(1-2p_{\rm err})^{r_i}$ to each term in the expectation value. Since the attenuator is an exponentially decreasing function in the weight $r_i$, only low--weight Pauli monomials contribute to the expectation value  $\langle{O}\rangle_M$. Thus, the proposed method is robust against readout error when estimating expectation values that have contributions dominantly from low--weight Pauli monomials (this does not necessarily exclude observables, such as projectors, that have contributions from all Pauli monomials: the contribution of higher weight Pauli monomials could be small on a class of states of interest, as in the case of the low entanglement states of this section's random projectors experiment). Finally, we remark that other errors, e.g. gate operation errors, could be included in this framework in a similar fashion and that, clearly, \eqref{eq:readout_err} is easily generalisable to the case of different readout errors per qubit.

\section{Related work} \label{Related work}
\textit{Quantum state tomography.}~In general, estimating an $N$--qubit state  to error $\epsilon$ in trace distance requires  ${\cal O}(2^N/\epsilon^2)$ copies of the state~\cite{ODonnell2016Efficient,Haah2017Sample}. The estimation of the density matrix with our tomographic procedure does not break this lower bound. If we choose the basis of $N$--qubit Pauli operators, Eq.~\eqref{general variance} tells us that a few matrix elements will require $M$ to be exponentially large in $N$. A different basis, such as the  standard basis $\ket{i}\bra{j}$, has exponentially many matrix elements bounded by 1. However, calculating the trace distance requires adding exponentially many of them, creating again the need for an exponentially large $M$. The key point of this paper, however, is to be able to estimate approximately and efficiently expectation values without resorting to the notion of density matrix.

\textit{Shadow tomography.}~In shadow tomography the task is to estimate with high probability the expectation values of a set of $K$ observables, to a certain precision $\epsilon$. It was shown that~\cite{Aaronson2018Shadow,Aaronson2019Gentle} this can be done using $M={\cal O}(\log(K)^2 N^2\epsilon^{-8})$ copies of the state, by implementing a non--local measurement on all copies of the state. Recently, these results were improved~\cite{Huang2020Predicting}. The authors of~\cite{Huang2020Predicting} proposed an algorithm where  $M={\cal O} (B\log(K)/\epsilon^2)$ copies of the state, with $B\geqslant\max_i \Vert{O_i}\Vert^2_{\rm shadow}$,  allow one to predict the expectation values of $K$ observables, $\{O_i\}_{i=1}^K$, to a precision $\epsilon$. The norm $\Vert{\cdot}\Vert_{\rm shadow}$ depends on the particular measurement procedure used in the algorithm. For example, for random Pauli measurements the authors show (Proposition~3 there) that for a $k$-local observable  $\Vert{O}\Vert_{\rm shadow}\leqslant2^k\Vert{O}\Vert_{\infty}$, where $\Vert\cdot\Vert_{\infty}$ denotes the spectral norm. 

The algorithm of~\cite{Huang2020Predicting} has a similar structure to the algorithm proposed in this work. Similarly to~\cite{Huang2020Predicting}, the choice of randomised single--qubit measurement directions of Eq.~\eqref{tomo N-qubit} is not the only possible one. The framework of~\cite{Paini2000Quantum,DAriano2003Spin} generates valid formulas for any irreducible projective representation in the space of the system, including for example the Clifford group and the representation of $SU(2)$ in the space of a spin $s=(2^N-1)/2$. Furthermore, one could also easily derive a general expression for the norm, bound of the standard deviation of the estimator, as $\sqrt{\tr[O^2]}$ multiplied by a factor only dependent on the group representation chosen. It is important to note, however, that the use of the Cauchy-Schwarz inequality in the derivation would generate suboptimal norms, often exponentially larger than the true standard deviation of the estimator. A tight norm (or seminorm) is key, as it gives a guarantee of performance with a lower number of preparations, especially if the separation from a suboptimal one is exponential. The closeness to the standard deviation of the estimator is what justifies the importance of  $\Vert{O}\Vert_2$ of~\eqref{original norm}: $\Vert{O}\Vert_2$ can be exponentially smaller than norms depending exponentially on the locality or on the dimension of the system (App.~\ref{app:pps} for the case of projectors onto the computational basis).

Finally, while the task in shadow tomography is to estimate with high probability the expectation values of a set of $K$ observables, to a certain precision $\epsilon$, we focus here on the estimation with high probability the expectation values of a given, yet arbitrary, observable to a certain precision $\epsilon$. We show that accomplishing this task requires $M={\cal O} (\Vert{O}\Vert^2/\epsilon^2)$ copies of the state. If a certain precision is required for the combined evaluation of expectation values of several observables, one would simply calculate the joint probability of several normal variables of being smaller than a certain value which would lead to a log contribution to the error in the number of observables.

\section{Conclusions and outlook} \label{Conclusions and outlook}
We have introduced a concept of approximate description of a quantum state for a system of $N$ qubits. We have shown that the approximate state comes with an explicit operational procedure to estimate the expectation value of any observable that admits an efficient representation in the Pauli basis with statistical error only dependent on the cardinality of the approximate quantum state and on the observable's seminorm, independent of $N$ and growing at most exponentially with the locality of the observable. The operational procedure to obtain the approximate state, other than the initial state preparations, only requires single--qubit operations and measurement. The natural robustness of random single--qubit operations to experimental noise, as we saw in the experimental results and also discussed in~\cite{Vermersch2019Probing}, suggests that the proposed protocol could be appropriate for NISQ computers.

We  expect the notion of the approximate quantum state to enable improvement over existing variational optimisation methods, as it would not require separate iterations for each set of values of the parameters. Machine learning optimisations, in particular for generative models known as Born Machines~\cite{Bornoriginal}, may be an interesting case, since the explicit dependence of the expectation values of the projectors onto computational basis states on the variational parameters~\cite{Bornmarcello} may be obtained even for a large number of qubits, since, as shown in App.~\ref{app N qubits}, the estimators for the projectors onto computational basis state have, with high probability, variance bounded by 1 for all values of $N$.

The definition of the snapshots, constituting an approximate quantum state, can be generalised, too. The snapshots can in fact be derived from a group different from $SU(2)$. This generalisation has interest that goes beyond the purely mathematical exercise. As we have seen, $SU(2)$ tomography works well for observables expressed as linear combinations of products of Pauli operators, especially if the weights of the Pauli monomials are low. Electronic Hamiltonians can be represented as linear combinations of products of Pauli operators, but at the cost, in general, of introducing Pauli monomials with weights increasing with $N$. In second quantisation, electronic Hamiltonians contain linear combinations of products of only two or four ladder operators for any $N$. The bosonic tomographic formulas of~\cite{Paini2000Quantum} can be extended to the fermionic case, choosing the additive group of Grassmann numbers~\cite{fermions}, instead of the additive group of complex numbers, as tomographic group and expressing the snapshots as vectors of sampled real numbers descending from Grassmann numbers and of measured presence or absence of electrons in orbitals. The estimators for ladder operators and for their linear combinations can then be calculated directly without the need for the mapping to Pauli operators and without the consequent appearance of weights of Pauli monomials growing with $N$.

\section{Acknowledgments}

MP would like to thank Chad Rigetti and Michael Brett for the support,  Amy Brown, Colm Ryan, David Garvin, Duncan Fletcher, Eric Peterson, Genya Crossman, John Lapeyre, John Macaulay, Juan Bello-Rivas, Mark Hodson, Max Henderson, Riccardo Manenti, Robert Smith and Sohaib Alam for the reviews and the feedback. AK acknowledges the support from the National Science Foundation  Grant No.~2037301.

\bibliographystyle{unsrtnat}
\bibliography{biblio}

\onecolumn
\appendix
\section{One qubit} \label{app 1 qubit}
Based on group-theoretical considerations,  it was shown in Refs.~\cite{Paini2000Quantum,DAriano2003Spin}  that the density operator on a  Hilbert space of spin-$s$ particle (a $2s+1$ dimensional space) can be written as
\begin{align}
\rho=\sum_{m_s=-s}^s\int_\Sigma\frac{d\vec n}{4\pi}p(m_s,\vec n) K(m_s,\vec n),        
\end{align}
with the unitary spherical surface ${\Sigma}$ as integration domain, $\vec n=(\cos \varphi\sin\vartheta,\sin\varphi\sin\vartheta,\cos\vartheta)$, with $\vartheta\in[0,\pi],\varphi\in[0,2\pi)$, $p(m,\vec n)$  probability of obtaining the result $m_s\in[-s,s]$ when measuring the spin in the eigenbasis of $\vec s\cdot\vec n$, where $\vec s=(s_x,s_y,s_z)$ is spin of the system, and $K(m_s,\vec n)$ is a kernel function given by
\begin{align}
K(m_s,\vec n)=\frac{2s+1}{\pi}\int_0^{2\pi}d\psi \sin^2\frac{\psi}{2} e^{i\psi (m_s-\vec s\cdot\vec n)}.  \label{kernel}       
\end{align}

Specialising Eq.~(\ref{kernel}) for a spin-$1/2$ particle 
where $m_s=\pm1/2\doteq m/2$  and $\vec{s}=(\sigma_x,\sigma_y,\sigma_z)/2\doteq\vec{\sigma}/2$  we can write
\begin{equation}
K_1(m,\vec{n})=\frac{2}{\pi}\int_0^{2\pi}d\psi\sin^2\frac\psi2 e^{i\frac{\psi}{2}(m-\vec\sigma\cdot\vec n)}, \label{k1 with pauli}
\end{equation}
where the subscript of $K_1$ is used to emphasise that is the kernel operator on a single qubit.
We calculate (\ref{k1 with pauli}) utilising the identity $e^{- i\frac{\psi}{2}\vec{\sigma}\cdot\vec{n}}=\mathbb{1}\cos{\frac{\psi}{2}}- i\,\vec{\sigma}\cdot\vec{n}\,\sin{\frac{\psi}{2}}$ and directly obtain
\begin{equation}\label{eq:k1}
K_1(m,\vec{n})=\frac{1}{2}(\mathbb{1}+ 3\,m\,\vec{\sigma}\cdot\vec{n}).
\end{equation}

In the main text we have defined
\begin{equation}
R_1[\sigma_\alpha](m,\vec{n})\doteq\tr[\sigma_\alpha K_1(m,\vec{n})]=3\,m\,n_\alpha, \label{R1 for pauli}
\end{equation}
an unbiased estimator of $\sigma_\alpha$, with $\alpha=x,y,z$ and $n_\alpha$ defined by $\vec{n}=(n_x,n_y,n_z)$. The  variance of $R_1[\sigma_\alpha]$ is 
\begin{align}
\var(R_1[\sigma_\alpha](m,\vec{n}))=\langle {R}_1^2[\sigma_\alpha](m,\vec n) \rangle-\langle {R}_1[\sigma_\alpha](m,\vec n) \rangle^2\leqslant \langle {R}_1^2[\sigma_\alpha](m,\vec n) \rangle. \label{variance defined}
\end{align}
We  calculate the last term of~\eqref{variance defined} utilising~\eqref{R1 for pauli}:
\begin{align}\label{variance of pauli}
\langle {R}_1[\sigma_\alpha](m,\vec n){R}_1[\sigma_\beta](m,\vec n) \rangle&=\int_{\Sigma}\frac{d\vec n}{4\pi}\sum_{m=\pm1}p(m,\vec{n}){R}_1[\sigma_\alpha](m,\vec{n}){R}_1[\sigma_\beta](m,\vec{n})\\\nonumber
&=9\int_{\Sigma}\frac{d\vec n}{4\pi}\,n_\alpha\,n_\beta \sum_{m=\pm1}p(m,\vec{n})=9\int_{\Sigma}\frac{d\vec n}{4\pi}\,n_\alpha\, n_\beta=3\,\delta_{\alpha\beta},
\end{align}
with $\delta_{\alpha\beta}$ is  Kronecker delta. To calculate the variance of $R_1[O](m,\vec{n})$, we expand $O$ in the Pauli operator basis $
O=\sum_{i=0}^{3}a_i\,\sigma_i$, so that
\begin{align}\label{main on variance}
\var(R_1[O](m,\vec{n}))&=\var\bigg(a_o\,R_1[\mathbb{1}](m,\vec{n}) + \sum_{i=1}^{3}a_i\,R_1[\sigma_i](m,\vec{n})\bigg)\\\nonumber
&=a_0^2\,\underbrace{\var(R_1[\mathbb{1}](m,\vec{n}))}_{=0}+\var\bigg(\sum_{i=1}^{3}a_i\,R_1[\sigma_i](m,\vec{n})\bigg)\\\nonumber
&+2\,\underbrace{\cov\bigg(R_1[\mathbb{1}](m,\vec{n}),\sum_{i=1}^{3}a_i\,R_1[\sigma_i](m,\vec{n})\bigg)}_{=0}\\\nonumber
&=\var\bigg(\sum_{i=1}^{3}a_i\,R_1[\sigma_i](m,\vec{n})\bigg)\leqslant\bigg\langle \bigg(\sum_{i=1}^{3}a_i\,R_1[\sigma_i](m,\vec{n})\bigg)^{\!2}\, \bigg\rangle\\\nonumber
&=\sum_{i=1}^{3}a_i^2\,\langle R_1^2[\sigma_i](m,\vec{n}) \rangle + \sum_{i\neq j (\neq 0)}\,a_i\,a_j \langle R_1[\sigma_i](m,\vec{n})R_1[\sigma_j](m,\vec{n})\rangle\\\nonumber
&=3\sum_{i=1}^{3}a_i^2+3 \sum_{i\neq j (\neq 0)}\,a_i\,a_j \,\delta_{ij}=3\sum_{i=1}^{3}a_i^2.
\end{align}
We define a seminorm  of $O$ as
\begin{align}
\Vert O\Vert=\sqrt{3\sum_{i=1}^{3}a_i^2} 										
\end{align}
and obtain 
\begin{align}
\var({R}_1[O](m,\vec n))\leqslant{\Vert O\Vert}^2.
\end{align}
Therefore, for $M$ sufficiently large, $\langle{O}\rangle_M$ has a normal statistical error smaller than or equal to $\Vert O\Vert/{\sqrt{M}}$.

\section{\boldmath{$N$} qubits} \label{app N qubits}
For a system of $N$ qubits, we expand the generic observable $O$ as in (\ref{general O})
\begin{align}
O=\sum_i \,a_i\,P_i										\label{general O again}
\end{align}
where $P_i=\sigma_{i_1}\otimes\cdots\otimes\sigma_{i_N}$,  $i=(i_1,\ldots\,i_N)$, is a Pauli monomial operator  and  $i_k=0,\ldots,3$,
and proceed in the calculation of the variance of $R[O]$ in a similar way to the one qubit case (the dependence of $R[\cdot]$ on $(m_1,\vec{n}_{1},\ldots,m_N,\vec{n}_{N})$ is not explicitly indicated for a lighter notation):
\begin{align}\label{main on variance N qubits}
\var(R[O])&=a_0^2\,\var(R[\mathbb{1}])+\var\bigg(\sum_{i\neq 0_v}a_i\,R[P_i]\bigg)+2\,\cov\bigg(R[\mathbb{1}],\sum_{i\neq 0_v}a_i\,R[P_i]\bigg)\\\nonumber
&=\var\bigg(\sum_{i\neq 0_v}a_i\,R[P_i]\bigg)\leqslant\bigg\langle \bigg(\sum_{i\neq 0_v}a_i\,R[P_i]\bigg)^{\!2}\, \bigg\rangle=\sum_{i,j (\neq 0_v)}\,a_i\,a_j \langle R[P_i]R[P_j]\rangle,
\end{align}
where the last sum on $i$ and $j$ is extended to all values of $i$ and $j$ except $0_v\doteq(0,0,\ldots,0)$. In general, for every $k$, the indices $i_k$ and $j_k$ appearing in each $\langle R[P_i]R[P_j]\rangle$ of (\ref{main on variance N qubits}) can assume any integer value between 0 and 3. For given $i$ and $j$, let $S$ be the set of indices $k$ for which $i_k\neq 0 \wedge  j_k\neq 0$, $s$ the cardinality of $S$ and $\overline S$ the set containing the remaining indices $k$ of ${1,\ldots,N}$ with cardinality ${\bar s}=N-s$. We use (\ref{R_N}) for a Pauli monomial to arrive at
\begin{align}\label{mother of all expansions}
\langle R[P_i]R[P_j]\rangle&=\bigg(\prod_{{h}\in{\overline S}}\int_{{\Sigma}_{h}}\frac{d{\vec n}_{h}}{4\pi}\sum_{m_{h}}{R}_1[\sigma_{i_{h}}](m_{i_{h}},\vec n_{i_{h}}){R}_1[\sigma_{j_{h}}](m_{j_{h}},\vec n_{j_{h}})\bigg)\\\nonumber
&\times\bigg(\prod_{k\in{S}}\int_{{\Sigma}_{k}}\frac{d{\vec n}_{k}}{4\pi}\sum_{m_{k}}{R}_1[\sigma_{i_{k}}](m_{i_{k}},\vec n_{i_{k}}){R}_1[\sigma_{j_{k}}](m_{j_{k}},\vec n_{j_{k}})\bigg)p(\{m,\vec n\})\\\nonumber
&=\bigg(\prod_{{h}\in{\overline S}}\int_{{\Sigma}_{h}}\frac{d{\vec n}_{h}}{4\pi}\sum_{m_{h}}{R}_1[\sigma_{i_{h}}](m_{i_{h}},\vec n_{i_{h}}){R}_1[\sigma_{j_{h}}](m_{j_{h}},\vec n_{j_{h}})\bigg)\\\nonumber
&\times\bigg(9^s\prod_{k\in{S}}\int_{{\Sigma}_{k}}\frac{d{\vec n}_{k}}{4\pi}\,{n}_{i_k}{n}_{j_k}\bigg)\bigg(\prod_{k\in{S}}
\sum_{m_{k}}p(\{m,\vec n\})\bigg)\\\nonumber
&=\bigg(\prod_{{h}\in{\overline S}}\int_{{\Sigma}_{h}}\frac{d{\vec n}_{h}}{4\pi}\sum_{m_{h}}{R}_1[\sigma_{i_{h}}](m_{i_{h}},\vec n_{i_{h}}){R}_1[\sigma_{j_{h}}](m_{j_{h}},\vec n_{j_{h}})\bigg)\\\nonumber
&\times\bigg(3^s \prod_{k\in{S}}\delta_{{i_k}{j_k}}\bigg)p(m_{h_1},\vec{n}_{h_1},\ldots,m_{h_{\bar{s}}},\vec{n}_{h_{\bar{s}}})\\\nonumber
&=\bigg(\prod_{{h}\in{\overline S}}\int_{{\Sigma}_{h}}\frac{d{\vec n}_{h}}{4\pi}\sum_{m_{h}}{R}_1[\sigma_{i_{h}}](m_{i_{h}},\vec n_{i_{h}}){R}_1[\sigma_{j_{h}}](m_{j_{h}},\vec n_{j_{h}})\bigg)\\\nonumber
&\times\bigg(3^s\prod_{k\in{S}}\delta_{{i_k}{j_k}}\int_{{\Sigma}_{k}}\frac{d{\vec n}_{k}}{4\pi}\sum_{m_{k}}{R}_1[\sigma_{i_{k}}^2]\bigg)p(\{m,\vec n\})\\\nonumber
&=3^s \bigg( \prod_{k\in{S}}\delta_{{i_k}{j_k}}\bigg)\bigg(\prod_{{h}\in{\overline S}}\int_{{\Sigma}_{h}}\frac{d{\vec n}_{h}}{4\pi}\sum_{m_{h}}{R}_1[\sigma_{i_{h}}\sigma_{j_{h}}](m_{i_{h}},\vec n_{i_{h}},m_{j_{h}},\vec n_{j_{h}})\bigg)\\\nonumber
&\times\bigg(\prod_{k\in{S}}\int_{{\Sigma}_{k}}\frac{d{\vec n}_{k}}{4\pi}\sum_{m_{k}}{R}_1[\sigma_{i_{k}}\sigma_{j_{k}}](m_{i_{k}},\vec n_{i_{k}},m_{j_{k}},\vec n_{j_{k}})\bigg)p(\{m,\vec n\})\\\nonumber
&=3^s \bigg( \prod_{k\in{S}}\delta_{{i_k}{j_k}}\bigg)\langle R[P_{i}P_{j}]\rangle=3^s \bigg( \prod_{k\in{S}}\delta_{{i_k}{j_k}}\bigg)\tr[P_{i}P_{j}\rho]\leqslant3^s \bigg( \prod_{k\in{S}}\delta_{{i_k}{j_k}}\bigg)\doteq3^{r_{ij}}\Delta_{ij}.\nonumber
\end{align}
If we introduce a seminorm of $O$ as
\begin{align}
\Vert O\Vert=\sqrt{\sum_{i,j (\neq 0_v)}3^{r_{ij}}\Delta_{ij}|a_i||a_j|}\,,
 										\label{norm N qubits again}
\end{align}
Eq. (\ref{main on variance N qubits}) becomes
\begin{align}
\var({R}[O](\{m,\vec n\}))\leqslant{\Vert O\Vert}^2.						
\end{align}

\par
\vskip 0.75em

The seminorm (\ref{norm N qubits again}) is a tighter upper bound for the variance of $R[O]$ than the formally simpler seminorm
\begin{align}
{\Vert O\Vert}_1=\sum_{i\neq 0_v}\sqrt{3^{r_{i}}}|a_i|,
 										\label{norm 1}
\end{align}
with $r_i$ representing the weight of $P_i$, since
\begin{align}\label{norm 1 inequality}
\Vert O\Vert^2=\sum_{i,j (\neq 0_v)}3^{r_{ij}}\Delta_{ij}|a_i||a_j|\leqslant\sum_{i,j (\neq 0_v)}\sqrt{ 3^{r_{i}} 3^{r_{j}}}|a_i||a_j|=\Bigl(\sum_{i\neq 0_v}\sqrt{ 3^{r_{i}}}|a_i|\Bigr)^2={\Vert O\Vert}_1^2
\end{align}
This is not surprising, since the seminorm (\ref{norm 1}) can be shown to be a bound for the variance of $R[O]$ by expanding $\var\bigl(\sum_{i\neq 0_v}a_i\,R[P_i]\bigr)$ of (\ref{main on variance N qubits}) and using Schwarz inequality without resorting to explicit properties of $R$ as in (\ref{mother of all expansions}). The seminorm (\ref{norm N qubits again}) can also be expressed as
\begin{align}
\Vert O\Vert^2={\Vert O\Vert}_2^2+\sum_{i\neq j (\neq 0_v)}3^{r_{ij}}\Delta_{ij}|a_i||a_j|,			\label{norm decomposition}
\end{align}
with the seminorm ${\Vert O\Vert}_2$ defined by
\begin{align}
{\Vert O\Vert}_2=\sqrt{\sum_{i\neq 0_v}3^{r_{i}}\,a_i^2}. 						\label{original norm N qubits}
\end{align}
We use (\ref{mother of all expansions}) and rewrite (\ref{main on variance N qubits}) as
\begin{align}
\var(R[O])\leqslant\sum_{i,j (\neq 0_v)}\,3^{r_{ij}}\,\Delta_{ij}\,\tr[P_{i}P_{j}\rho]\,a_i\,a_j={\Vert O\Vert}_2^2+\sum_{i\neq j (\neq 0_v)}\,3^{r_{ij}}\,\Delta_{ij}\,\tr[P_{i}P_{j}\rho]\,a_i\,a_j.		\label{variance and mixed terms}
\end{align}
Eq.~\eqref{variance and mixed terms} shows how the mixed terms of (\ref{norm decomposition}) are a weak upper bound for the last term of (\ref{variance and mixed terms}). For a randomly chosen $O$ or, in the same way, for a random $\rho$, the addends in the last term of (\ref{variance and mixed terms}) have mixed positive and negative signs. Furthermore, for most states $\rho$, the expectation values of the expressions $P_{i}P_{j}$ tend to become smaller for higher values of $r_{ij}$. As a result, ${\Vert O\Vert}_2$ is generally a good approximation of the standard deviation of $R[O]$.

To formalise this statement, we start by showing that the expectation value of the mixed terms of (\ref{variance and mixed terms}) over the space of the density operators of the system is equal to zero. The probability of choosing a density operator from the space of density operators is the probability of choosing a set of eigenvalues multiplied by the probability of choosing a density operator with specific eigenvalues conditional to the choice of the eigenvalues. If for any choice of the eigenvalues, the expectation value over the density operators with the chosen eigenvalues is zero, then the expectation value over all density operators is zero. All density operators with the same eigenvalues can be obtained from one with a unitary transformation and, vice versa, all unitary transformations applied to a density operator preserve the eigenvalues. Therefore, we can calculate the expectation value of the mixed terms of (\ref{variance and mixed terms}) over the density operators with given eigenvalues by averaging over the unitary group~\cite{Li2013SelectedTI} with the unitary transformations applied to an arbitrary density operator $\rho_0$
\begin{align}\label{unitary average}
&\int_U dU\sum_{i\neq j (\neq 0_v)}3^{r_{ij}}\,\Delta_{ij}\,a_i\,a_j\,\tr[P_{i}P_{j}\,U\rho_0\,U^{\dag}]
=\sum_{i\neq j (\neq 0_v)}3^{r_{ij}}\,\Delta_{ij}\,a_i\,a_j\,\tr\bigg[P_{i}P_{j}\int_U dU\, U\rho_0\,U^{\dag}\bigg]\\
&=\sum_{i\neq j (\neq 0_v)}3^{r_{ij}}\,\Delta_{ij}\,a_i\,a_j\,\tr\Bigl[\frac{P_{i}P_{j}}{2^N}\Bigr]=0,\nonumber
\end{align}
which implies
\begin{align}\label{overall average}
\bigg\langle \sum_{i\neq j (\neq 0_v)}\,3^{r_{ij}}\,\Delta_{ij}\,\tr[P_{i}P_{j}\rho]\,a_i\,a_j\bigg\rangle_{\!\!\rho}=0.
\end{align}
To gain further information on how close ${\Vert O\Vert}_2$ is likely to be to ${\Vert O\Vert}$ for a random state, we now consider the variance of the mixed terms of (\ref{norm decomposition}) over the space of the density operators of the system. For simplicity, we specialise the analysis to pure states. From the unitary group notation of (\ref{unitary average}), we switch to the average over the states expressed as in~\cite{Ambainis:2007:QTT:1251970.1252159} and recall the following identities:
\begin{align}
\int_\psi{d\psi}\ket{\psi}\!\bra{\psi}^{\otimes 2}=\frac2{2^N(2^N+1)}\,\mathbb{1}_{\rm sym},
\end{align}
with $\mathbb{1}_{\rm sym}$ projector onto the symmetric subspace of ${\cal H}^{\otimes 2}$, and
\begin{align}
\tr[A\otimes B\,\mathbb{1}_{\rm sym}]=\frac1{2}\big(\tr[A]\tr[B]+\tr[AB]\big),
\end{align}
for any operator $A$ and $B$. The variance of the mixed terms of (\ref{variance and mixed terms}) over the states of the system can be expressed as
\begin{align}\label{variance of mixed terms}
&\var\bigg(\sum_{i\neq j (\neq 0_v)}\,3^{r_{ij}}\,\Delta_{ij}\,\tr\big[P_{i}P_{j}\ket{\psi}\!\bra{\psi}\big]\,a_i\,a_j\bigg)_{\!\psi}=\int_\psi{d\psi}\bigg(\sum_{i\neq j(\neq 0_v)}a_i\,a_j\,3^{r_{ij}}\,\Delta_{ij}\,\tr\big[P_{i}P_{j}\ket{\psi}\!\bra{\psi}\big]\bigg)^{\! 2}\\\nonumber
&=\sum_{i\neq j(\neq 0_v)}\sum_{i'\neq j'(\neq 0_v)}a_i\,a_j\,a_{i'}a_{j'}\,3^{r_{ij}}\,3^{r_{i'j'}}\,\Delta_{ij}\,\Delta_{i'j'}\int_\psi{d\psi}\,\big(\tr\big[P_{i}P_{j}\ket{\psi}\!\bra{\psi}\big]\,\tr\big[P_{i'}P_{j'}\ket{\psi}\!\bra{\psi}\big]\big)\\\nonumber
&=\sum_{i\neq j(\neq 0_v)}\sum_{i'\neq j'(\neq 0_v)}a_i\,a_j\,a_{i'}a_{j'}\,3^{r_{ij}}\,3^{r_{i'j'}}\,\Delta_{ij}\,\Delta_{i'j'}\,\tr\bigg[(P_{i}P_{j})\otimes (P_{i'}P_{j'})\int_\psi{d\psi}\ket{\psi}\!\bra{\psi}^{\otimes 2}\bigg]\\\nonumber
&=\frac2{2^N(2^N+1)}\sum_{i\neq j(\neq 0_v)}\sum_{i'\neq j'(\neq 0_v)}a_i\,a_j\,a_{i'}a_{j'}\,3^{r_{ij}}\,3^{r_{i'j'}}\,\Delta_{ij}\,\Delta_{i'j'}\,\tr\big[(P_{i}P_{j})\otimes (P_{i'}P_{j'})\,\mathbb{1}_{\rm sym}\big]\\\nonumber
&=\frac1{2^N(2^N+1)}\sum_{i\neq j(\neq 0_v)}\sum_{i'\neq j'(\neq 0_v)}a_i\,a_j\,a_{i'}a_{j'}\,3^{r_{ij}}\,3^{r_{i'j'}}\,\Delta_{ij}\,\Delta_{i'j'}\,\tr[P_{i}P_{j}P_{i'}P_{j'}]\\\nonumber
&=\frac1{(2^N+1)}\sum_{i\neq j(\neq 0_v)}\sum_{i'\neq j'(\neq 0_v)}a_i\,a_j\,a_{i'}a_{j'}\,3^{r_{ij}}\,3^{r_{i'j'}}\,\Delta_{ij}\,\Delta_{i'j'}\,\overline{\Delta}_{iji'j'},\\\nonumber
\end{align}
with $\overline{\Delta}_{iji'j'}=1$ when $i,j,i',j'$ are such that $P_{i}P_{j}P_{i'}P_{j'}=\mathbb{1}$, otherwise $\overline{\Delta}_{iji'j'}=0$. A general upper bound of the last expression is likely to be weak, given its strong dependence on $O$. The specific case of $O$ given by a projector onto a computational basis state is examined in App.~\ref{app:pps}.

\section{Projectors onto the computational basis states} \label{app:pps}
In this section we examine properties of the seminorm for an important special case, namely, the case of projectors onto a generic element of the computational basis state,
\begin{align}
\ket{x}\bra{x}=\bigotimes_{k=1}^N\,\frac{1}{2} (\mathbb{1}\pm\sigma_{z_k}),    \label{projector}
\end{align}
where the $\pm$ of each factor is defined by the choice of the element of the computational basis $\ket{x}$. This case is important not only for its practical significance in a machine learning context, as briefly indicated in Sec.~\ref{Conclusions and outlook}, but also because it is representative of the cases with bounded ${\Vert O\Vert}_2$ and diverging ${\Vert O\Vert}$ for increasing $N$,  which makes the determination of whether the statistical error of the estimator can be approximated by ${\Vert O\Vert}_2$ of particular interest. To calculate the seminorms of any projector onto a computational basis state, we expand the tensor product of (\ref{projector}) and use the seminorm definitions. For ${\Vert \ket{x}\bra{x}\Vert}_2$ this simply means
\begin{align}
{\Vert \ket{x}\bra{x}\Vert}_2^2=\frac{1}{4^N}\bigg({{N}\choose{N-1}}\,3^1+{{N}\choose{N-2}}\,3^2+\ldots+{{N}\choose{0}}\,3^N\bigg)=\frac{1}{4^N}\big((1+3)^N\big)-\frac{1}{4^N}=1-\frac{1}{4^N}.
    \label{norm_2 projector}
\end{align}
A more involved combinatorial calculation would show that ${\Vert \ket{x}\bra{x}\Vert}^2\leqslant(3/2)^N$ and ${\Vert \ket{x}\bra{x}\Vert}^2=O((3/2)^N)$. We come back to (\ref{variance of mixed terms}) for any projector onto computational basis state and observe that, for given $i$ and $j$, there are $2^N$ ordered pairs $i',j'$ such that $\overline{\Delta}_{iji'j'}=1$. The operator $P_{i}P_{j}$ is in fact the tensor product of single--qubit identities and operators $\sigma_z$. Each of them becomes the identity when multiplied by itself and this can be obtained with two choices for every qubit of $P_{i'}$ and $P_{j'}$. Unfortunately, to complicate things, ${r_{i'j'}}$ is not independent of the choice of $i'$ and $j'$ for given $i$ and $j$; however, we can simplify the evaluation of (\ref{variance of mixed terms}) by considering a limit case and determining an upper bound. In particular, for every $i$ and $j$, with $i\neq j$ and $i$ and $j$ different from $0_v$,
\begin{align}    \label{variance of mixed terms of projectors}
\sum_{i',j'}\,3^{r_{i'j'}}\overline{\Delta}_{iji'j'}<\sum_{i',j'}\,3^{r_{i'j'}}\overline{\Delta}_{iii'j'}=\sum_{l=0}^N{{N}\choose{l}}\,3^{l}=4^N.
\end{align}
With the substitutions $a_i=(1/2)^N$ and $\Delta_{ij}=1$ valid for any projector $\ket{x}\bra{x}$ and any value of the indexes, the last term of Eq. (\ref{variance of mixed terms}) becomes
\begin{align}\label{smaller than 3/4}
&\frac1{(2^N+1)}\sum_{i\neq j(\neq 0_v)}\sum_{i'\neq j'(\neq 0_v)}a_i\,a_j\,a_{i'}a_{j'}\,3^{r_{ij}}\,3^{r_{i'j'}}\,\Delta_{ij}\,\Delta_{i'j'}\,\overline{\Delta}_{iji'j'}
\\\nonumber
&<\frac{1}{(2^N+1)\,4^N}\sum_{i\neq j(\neq 0_v)}3^{r_{ij}}=\frac{{\Vert \ket{x}\bra{x}\Vert}^2}{2^N+1}<\bigg(\frac{3}{4}\bigg)^{\!N}.
\end{align}
Eq.~\eqref{smaller than 3/4} shows that, for any projector $\ket{x}\bra{x}$, $\var(R[\ket{x}\bra{x}])$ for a random state is bounded by ${\Vert \ket{x}\bra{x}\Vert}_2^2<1$ with probability exponentially close to 1 as a function of $N$.

\end{document}